# Cavity-Enhanced Raman Scattering from 2D Hybrid Perovskites

Aditya Singh[1,2], Jason Lynch[2], Surendra B. Anantharaman[2], Jin Hou[3], Simrjit Singh[2], Gwangwoo Kim[2], Aditya D. Mohite[3], Rajendra Singh[1], and Deep Jariwala[2*]

[1]Department of Physics, Indian Institute of Technology Delhi, New Delhi-110016, India

[2]Department of Electrical and Systems Engineering, University of Pennsylvania, Philadelphia, Pennsylvania 19104, United States

[3]Department of Chemical and Biomolecular Engineering and Department of Materials Science and Nanoengineering, Rice University, Houston, Texas 77005, United States

*Corresponding author: dmj@seas.upenn.edu

# ABSTRACT

Two-dimensional (2D) hybrid organic-inorganic perovskites (HOIPs) are promising candidates for optoelectronic applications due to their efficient light emission properties, and strong dielectric confinement effects. Raman spectroscopy is a versatile, non-contact and often non-destructive technique, widely used to characterize crystalline materials. However, the inherently weak phonon scattering, strong background, and complex nature of the Raman signals from HOIPs present challenges in obtaining reliable signals. Further, the fragile nature of the 2D HOIP crystals results in rapid degradation upon exposure to heat, light and moisture, which presents further difficulty in enhancing Raman scattered photon signals. Herein, we report a novel approach to enhance the weak Raman scattering signals in Ruddlesden-Popper (RP) phase HOIPs by introducing an open-cavity comprising HOIP crystals on a gold substrate. We observe 15x enhancement of the Raman signals due to the Purcell effect inside the high-index HOIP crystals. Our simple approach can be extended to enhance the study of phonon scattering in other HOIP and van der Waals layered crystals.

## 1. INTRODUCTION

Two-dimensional (2D) Ruddlesden−Popper (RP) hybrid organic-inorganic perovskites (HOIPs) have attracted much attention due to their uniquely tunable structure, strong quantum confinement and extraordinary performance in optoelectronic devices[1]. The RP phase of these HOIPs have a layered crystal structure represented by the chemical formula $(RNH_3)_2(MNH_3)_{n-1}A_nX_{3n+1}$, where 'R' is an alkyl or aromatic moiety, 'M' is a methyl group (-$CH_3$), 'A' is a metal cation (Pb), 'X' is a halide (I, Cl, Br), and 'n' is an integer that defines the number of inorganic ($PbI_2$) layers of corner-sharing $[PbI_6]^{4-}$ octahedra confined by two adjacent organic layers[2]. $[PbI_6]^{4-}$ layers act as quantum wells, and organic layers as dielectric spacers which play a crucial role in charge confinement, hydrophobicity, structural rigidity, and stability[3].

Physical properties of HOIPs such as charge confinement can be tuned by the integer n, electron-phonon (e-p) coupling by organic moiety, and band structure by halide anion and external strain[3,4]. Further, the e-p coupling is responsible for scattering and provides decay channels for the excitons, which leads to the determination of absorption and emission properties of the material[5,6]. Therefore, investigation of vibrational dynamics of HOIPs is of paramount importance as they possess the information about e-p interaction, structural and compositional changes, symmetry of lattice vibration, temperature and pressure phase diagram, degree of crystallinity, dielectric screening, as well as elastic properties of the crystal[2,3].

Raman spectroscopy plays a crucial role in the study of vibrational dynamics of materials. Unfortunately, HOIPs have inherently weak optical phonon scattering and strong background from the relaxation of electronic excited states[7] when excited above the bandgap. In addition, several overlapping and complex Raman modes smother crucial information about their vibrational spectra and properties [8]. HOIPs exposed to excessive power of laser excitation within the bandgap of the material are prone to degradation due to the thermal stress induced by the laser[8–10]. Degradation in HOIP crystals introduces additional Raman peaks artifacts making the characterization of the material, and the assignment of the modes very complex and challenging [3,11,12]. These material challenges/limitations compel the search and development of a consistent technique or means to characterize the vibrational spectra from such environment sensitive materials via Raman scattering.

HOIPs have crystal lattice that intrinsically exhibits low probability for Raman scattering. Enhancement in Raman scattering can be achieved by electromagnetic enhancement by placing the semiconductor over metallic nanoparticles for surface-enhanced Raman spectroscopy, or placing them between nanogap metallic cavities for tip-enhanced Raman spectroscopy[13]. However, the aforementioned material limitations in terms of thermal and chemical stability constitute a significant drawback in enhancing the phonon scattering from the HOIP crystals.

In this paper, we fabricated an open-cavity comprising the RP perovskite flakes on a gold substrate (which acts as a reflector) and exploit the Purcell effect to increase the probability of Raman scattering by enhancing light trapping in the crystal. The extraordinary optical constants of HOIP crystals facilitate the strong light-matter coupling in thick crystals (thickness, t > 100 nm) which leads to photonic open-cavity formation in crystals themselves[1,14]. When the laser shines on the HOIP/gold, the laser undergoes multiple reflections inside the dielectric cavity created by the flat surfaces of the HOIP crystal. Due to the Purcell effect inside the crystals, the optical electric field is enhanced which results in enhanced Raman scattering by a factor of up to 15x. Our work exhibits a simple approach to enhance Raman scattering which can be extended to enhance phonon scattering in other van der Waals crystals.

## 2. EXPERIMENTAL METHODS

**Sample preparation:** Gold template substrates were prepared by depositing a 100 nm thick film of gold on a silicon wafer, and a mixer of two epoxy components was dropped over it. Silicon substrates were placed over epoxy drops and heated to 100 °C for 30 min. After cooling, template-stripped gold substrates were ready to use[15]. Hybrid organic-inorganic perovskite (HOIP) in Ruddlesden-Popper (RP) phase flakes were exfoliated from the single crystal samples by the scotch-tape method inside the glove box to avoid moisture and sample degradation.

**Optical characterization:** Raman scattering and reflectance measurements were performed by 600 grooves/mm grating and 50x objective (NA=0.35) in Horiba LabRam HR Evolution system under the vacuum ($10^{-5}$ torr) using Linkam stage. For low-temperature measurements, liquid nitrogen was used as a coolant in the Linkam stage. Raman scattering measurements were carried out using 633 nm and 785 nm laser wavelengths as excitation source. Reflectance measurements of samples were recorded by illuminating external white light source on the samples. Reflectance from the silver mirror was used to normalize the reflectance data and to avoid the effect of gold substrate absorption in the reflectance data of samples.

**Atomic force microscopy:** Thickness and surface topography of samples were measured by OmegaScope Smart SPM (AIST).

**Electric field distribution simulation:** Transfer-matrix-method (TMM) was implemented using python based on the theory presented in the reference[16] to simulate the electric field (E-field) distribution profiles of 633 and 785 nm lasers for RPn placed on quartz and gold substrates. Required optical constants of RPn have been obtained from ellipsometry measurements[14].

## 3. RESULTS AND DISCUSSION

Figure 1a shows the schematic of the crystal structure of RP phase lead iodide based HOIP whose general chemical formula is $(BA)_2(MA)_{n-1}Pb_nI_{3n+1}$, where BA is butylammonium and MA is methylammonium[2,3]. Therefore, RPn (n= 1, 2, 3…) can be chemically expressed as $(BA)_2PbI_4$, $(BA)_2(MA)Pb_2I_7$ and $(BA)_2(MA)_2Pb_3I_{10}$ or in short, RP1, RP2 and RP3, respectively. In our previous work[1], we reported that thick (t > 100 nm) RP HOIP crystals placed on gold show strong light-matter coupling (exciton-polariton formation) and form an open-cavity with quality factor ~250. Here, we exploit the Purcell effect in HOIP to enhance the emission of Raman scattering, using the 633 and 785 nm lasers, which are far from the excitonic absorption (Figure S1, Supporting Information). Purcell proposed that spontaneous emission rate can be modified (enhanced) by placing the emitter inside a cavity[17]. Using the Purcell effect, enhanced radiative emission from a variety of semiconducting and atomic systems has been reported[18,19]. Here, we use the same phenomenon for enhancing the Raman scattering signals from a weak phonon scattering crystal such as 2D HOIPs.

In our samples, light is trapped inside the cavity due to multireflection from the bottom gold substrate and top HOIP crystal-air interface. The required minimum volume of the cavity under diffraction limit is approximately calculated by

$$0.1 \left(\frac{\lambda_0}{n'}\right)^3 \qquad (1)$$

Where $\lambda_0$ is the free-space wavelength of light and n' is the refractive index of the material inside the cavity[20]. Refractive indices of RP1, RP2, and RP3 for $\lambda_0$ = 785 (633) nm wavelength was obtained from spectroscopic ellipsometry as 2.1 (2.2), 2.2 (2.4), and 2.3 (2.7), respectively[14]. So, from equation 1, it turns out that all three RPn achieve the minimum size/thickness of the cavity to trap light in crystals that are ≳ 100 nm at wavelengths of 633 and 785 nm, which are primary pump lasers for Raman scattering experiment. These theoretical calculations match well with our experimental results of reflectance, which show exciton-polariton formation in thick HOIP crystals (t > 100 nm) rather than thin ones (t < 30 nm) (Figure S1) as reported before [1].

Figure 1b depicts the interaction of the incident laser beam with HOIP crystal placed on a quartz substrate. Due to low reflectivity of the HOIP-quartz interface, most of the photons pass through it and do not reflect, limiting the number of reflections inside the cavity, due to which multiple reflections of the laser inside the crystal don't occur. Laser (incident photons) interacts with HOIP crystals and back-scattered signal photons collected by the objective are passed through the spectrometer to the detector. However, when the transmissive quartz substrate is replaced with a highly reflective gold substrate, the incident laser beam reflects backs at the HOIP-gold interface. Due to the cavity nature of 2D HOIP thick crystals (high refractive index), laser again reflects back from the top of the HOIP-air interface, which leads to multiple reflections inside the crystals. The incident and reflected laser beams undergo interference resulting in electric field enhancement in the crystal.

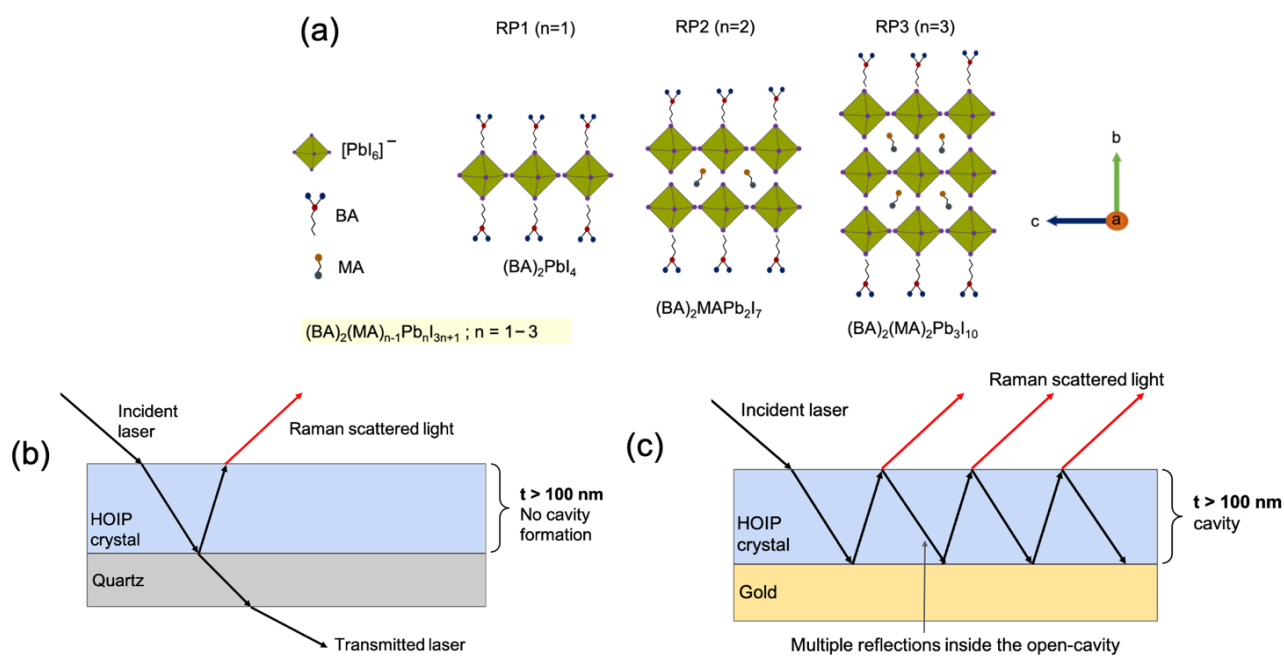

*Figure 1. Schematic illustration of enhanced Raman signals from 2D Ruddlesden−Popper (RP) hybrid organic-inorganic perovskites (HOIP) due to cavity formation. (a) Schematic of crystal structure of RP phase 2D hybrid butylamine (BA) lead iodide based perovskites, where n is the number of inorganic layers of [PbI$_6$]$^{4-}$ between two organic layers. (b) Schematic of the interaction of incident laser with thick (t > 100 nm) HOIP crystals placed over a quartz substrate. Laser (photons) interact with crystal, scattering occurs, and laser transmits through quartz substrate. (c) When quartz substrate is replaced by optically thick gold, the reflected laser beam interferes with the incoming beam. The black arrow shows the electric field propagation direction of the pump laser, while the red arrow shows the electric field propagation direction of the scattered. Note: the incident laser shown in this schematic is at an angle, while in Raman experiments/electric field simulation, the laser is at normal incidence over the crystal surface.*

Figures 2a,b show the Raman spectra of thick HOIP/quartz and HOIP/gold measured at 80 K using a 785 nm laser source. Information about Raman modes of inorganic-organic perovskite and organic spacer layers is enclosed in the range < 200 cm$^{-1}$ (range R1, hereafter) and 1200-1800 cm$^{-1}$ (range R2, hereafter)[21]. Details of frequency and character of these vibrational modes are explained in the text later. The thicknesses of these crystals have been measured by atomic force microscopy (AFM), and corresponding topographical images and height profiles are shown in Supporting Information (Figure S2). As 785 nm laser is used for pumping which is non-resonant with the exciton-polariton modes from all three RPn used in this study (Figure S1). Therefore, the enhanced Raman intensity is purely attributed to the cavity effect rather than resonant Raman scattering. As a result, thick crystals of RPn/gold show enhanced Raman intensity up to 15x compared to RPn/quartz due to cavity enhanced Raman scattering (see Figures 2a,b). In range R1, Raman mode (measured by 785 nm laser) of RPs at ≈ 48 cm$^{-1}$, which belongs to the PbI$_3$ network, has enhancement in the intensity of 3x, 15x, and 7x for RP1, RP2, and RP3, respectively (Figure 2a). However, modes at ≈ 92 and ≈ 103 cm$^{-1}$, and all modes of range R2 (for 785 and 633 lasers both), are very weak and poorly resolved in RPs/quartz compared to RPs/gold, which restricts the quantification of enhancement in Raman intensity. Further, in the case of Raman of RP1 and RP2 by 633 nm laser, there is an enhancement of 2-3x for various modes in range R1 (Figures 2c,e).

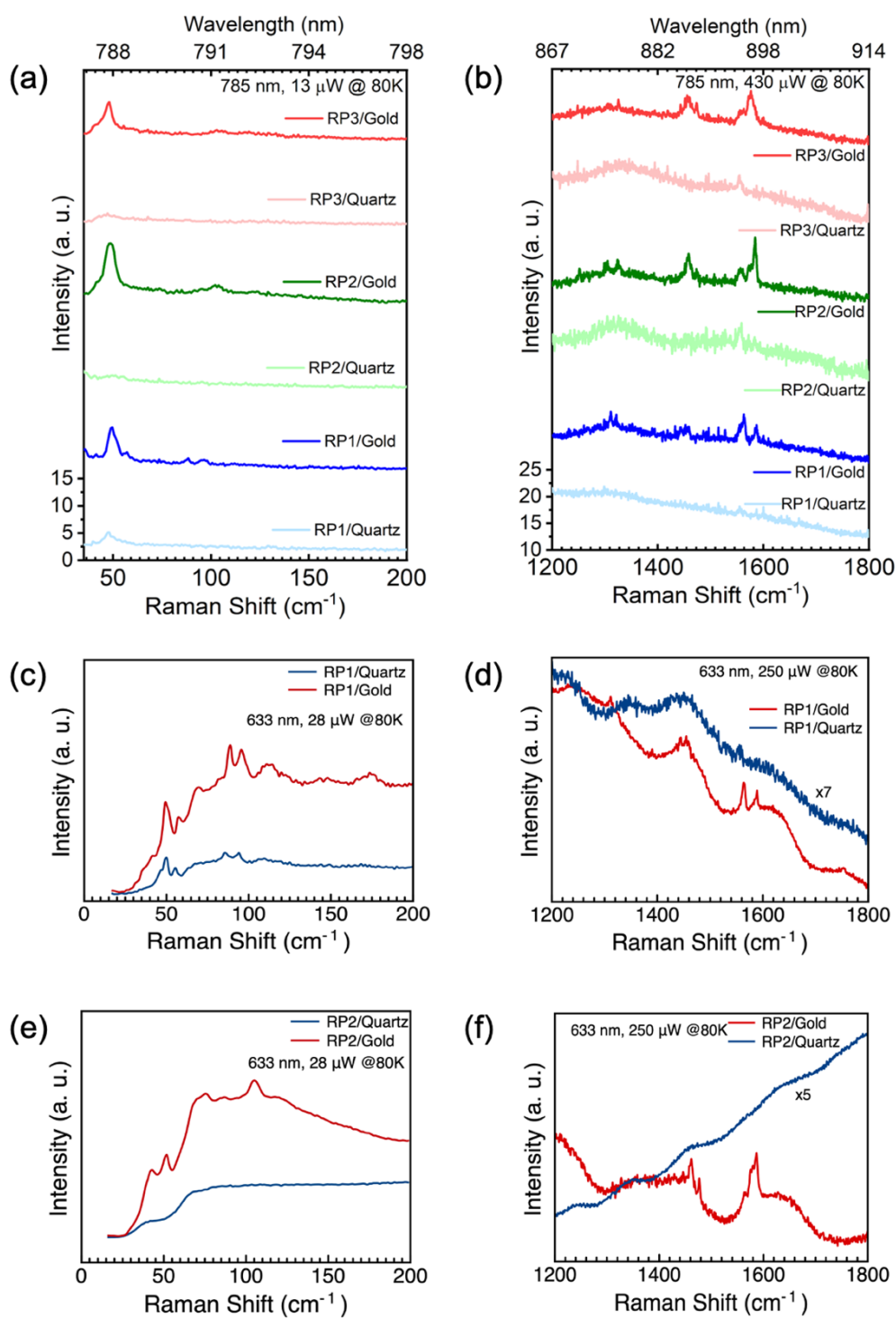

*Figure 2.* Cavity-enhanced Raman signals from thick crystals of RPn/gold compared to RPn/quartz measured by 785 nm laser line (non-resonant to all RPn) at 80K for (a) inorganic and organic layers and (b) organic layers of HOIP crystals. Raman spectra of RP1 (c, d) and RP2 (e, f) by 633 nm laser at 80K.

To validate the universality of the open-cavity effect, Raman spectra of the same samples were recorded using 633 nm excitation laser as well (Figures 2c-f). It is worth mentioning that, the huge background in Raman spectra by 633 nm laser, unlike the flat background by 785 nm, is due to our broad cut-off wavelength in the 633 nm edge filter of the Raman system (see Figure S3, Supporting Information). At 80K, Raman signals of thick RP3 by the 633 nm laser were subjugated by polaritonic emission at ~ 645 nm (see Figure S1, Supporting Information). Compared to the 785 nm laser, Raman measurements by the 633 nm laser show more well-resolved modes at low powers (< 30 µW) due to (a) Raman scattering intensity is inversely proportional to excitation wavelength and (b) sensitivity of charge-coupled device detector decreases in the near-infrared region, which applies for the 785 nm laser wavelength[22]. Further, we did Raman measurements of RP3/quartz by 633 nm laser source and found that although it leads to near-resonant Raman scattering, the cavity effect is stronger (Figure S5).

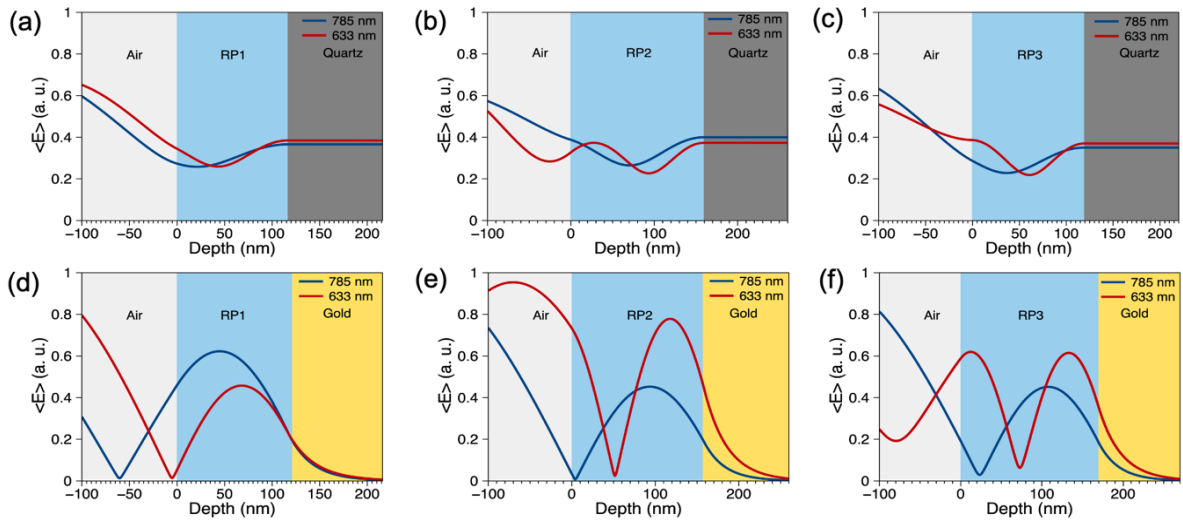

*Figure 3. Transfer-matrix-method simulations of the time-averaged electric field distribution of 633 and 785 nm lasers in thick (t > 100 nm) HOIP crystals:* E-field distribution profiles in thick crystals of RPn on (a-c) quartz and (d-f) gold substrate. The thickness of RP1, RP2, and RP3 crystal on quartz (gold) is 117 (120), 160 (157), and 120 (170) nm, respectively.

Figure 3 shows the time-averaged E-field distribution profile of 633 and 785 nm lasers in thick crystals (t > 100 nm) of RPn/quartz and RPn/gold. The AFM measured thicknesses of RP1, RP2, and RP3 crystals on quartz (gold) are 117 (120), 160 (157), and 120 (170) nanometers, respectively (Figure S2). RPn/quartz have minimum E-field intensity near the middle of the thickness of the crystal (Figures 3a-c). Further, the transparent nature of quartz makes laser beam pass through it leading to limited back reflection and E-field enhancement inside the crystal. However, thick HOIP crystals on gold have a maximum enhanced E-field in the crystal (Figures 3d-e). Higher optical E-fields within the crystal vis-a-vis the air-crystal and crystal-substrate interface increases the probability for interaction between the laser beam and the phonons of HOIP crystal, which ultimately leads to enhanced Raman intensity[20,23]. We have also verified that surface plasmons do not play any significant role in this Raman intensity enhancement via a separate control experiment (see supporting information S6).

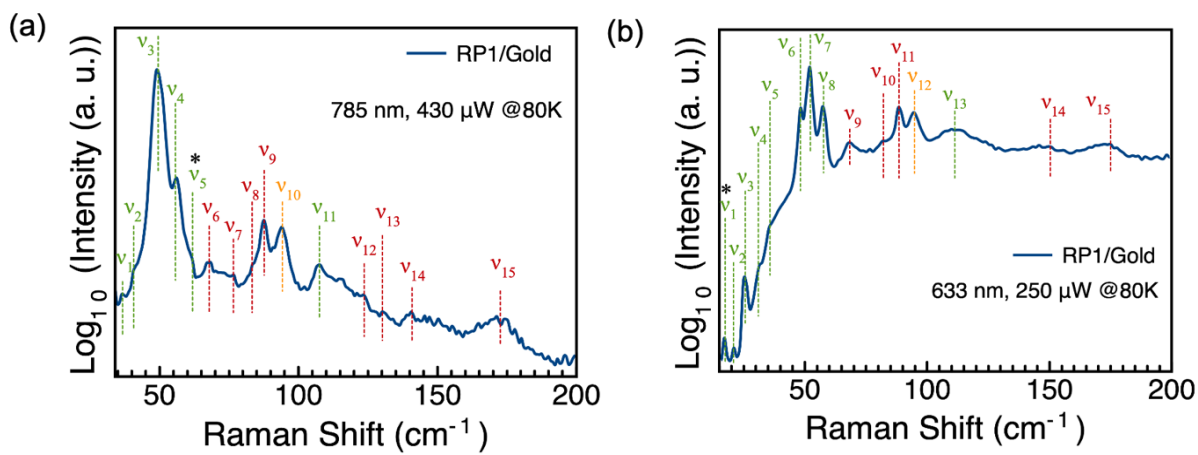

***Figure 4. Application of enhanced Raman intensity:*** *Logarithmic plot of Raman spectra of RP1/gold at 80K obtained via pumping (a) 785 and (b) 633 nm lasers. Raman modes in green and red belong to inorganic and organic components of the HOIPs, respectively, while modes in orange are common to both. Modes marked with asterisk (\*) are theoretically predicted to be silent but have been observed in our experiments.*

Lattice vibrations of HOIP crystals in Raman scattering spectra can be characterized notably into two sections: range R1 (< 200 cm$^{-1}$), which comprise the vibrational dynamics of the inorganic layer of PbI$_3$ network and organic layer both, representing low wavenumbers, and range R2 (1200-1800 cm$^{-1}$), which contains the lattice vibration of the organic layer only[21]. Recently, Menahem et al.[24] did Raman spectroscopy and density functional perturbation theory of 2D (BA)$_2$PbI$_4$ crystals for the range < 200 cm$^{-1}$ and compared their structural dynamics with its 3D counterpart, MAPbI$_3$. They concluded that like MAPbI$_3$, (BA)$_2$PbI$_4$ also exhibits temperature-dependent phase transition. At room-temperature, (BA)$_2$PbI$_4$ has a tetragonal crystal structure, while at low temperatures, it has an orthorhombic phase. Further, both materials have similar vibrational dynamics, and their Raman modes appear approximately at the same positions at 80K[21,24]. It is noteworthy that as organic spacers BA and MA have similar chemical nature. Therefore, we can assume that the Raman features of $(BA)_2PbI_4$, are similar to that of $MAPbI_3$.

A normal mode of vibration is whether Raman- or IR-active, it depends upon its symmetry[25]. Group theory analysis shows that the symmetries of the normal vibrational modes of $MAPbI_3$ can be expressed as: $\Gamma_{optic}$ = $17A_u$ + $21B_{1u}$ + $16B_{2u}$ + $21B_{3u}$ + $19A_g$ + $14B_{1g}$ + $19B_{2g}$ + $14B_{3g}$. In case of inorganic $PbI_3$ network, $A_u$ modes are expected to be silent, u modes are IR-active and g modes are Raman-active[26]. Figures 4a, b show the logarithmic plot of Raman spectra of RP1/gold at 80 K by 785 and 633 nm lasers, respectively. We observe a wide variety of Raman peaks in the R1 range from our cavity-enhanced Raman spectra. Raman peaks ($v_1$-$v_{15}$) measured by the 785 nm laser are observed at 36.4, 40.5, 48.8, 56.2, 61.8, 67.7, 76.0, 83.4, 87.5, 94.0, 107.3, 123.4, 130.0, 146.5, 171.3 cm$^{-1}$ and by the 633 nm laser are observed at 17.0, 21.2, 25.3, 31.2, 35.2, 48.5, 52.1, 57.0, 67.5, 81.7, 88.5, 94.5, 111.0, 150.0, 174.7 cm$^{-1}$. Raman modes < 65 cm$^{-1}$ are dominated by $PbI_3$ network only, which consist of rocking and bending vibrations of Pb-I-Pb, while in the range 65– 200 cm$^{-1}$, vibrations are either vibration/translation modes of the organic cations or internal vibrations of the $PbI_3$ network[21]. In Figures 4a and b, Raman modes in green and red belong to inorganic and organic layers, respectively, while the mode in orange overlaps with both. Modes marked with an asterisk (*) are theoretically silent ($A_u$ modes) but were observed in our experiments[21,24]. Further, in our enhanced Raman signals we also observe the silent $A_u$ modes at 17.0 and 61.8 cm$^{-1}$ in $(BA)_2PbI_4$ which have not been reported in literature precedent to the best of our knowledge. These theoretically silent peaks have been assigned to the Pb-I-Pb bend and Pb-I stretch modes in $MAPbI_3$, respectively[21].

Fundamental Raman modes of BA cation fall in the R2 range (1200 -1800 cm$^{-1}$) which comprises of the angular distortion of CH$_2$, CH$_3$, and NH$_2$ groups[27,28]. Observed Raman modes in our experiments in R2 range at ≈1251 cm$^{-1}$ correspond to CH$_2$ twisting, ≈1300 cm$^{-1}$ to CH$_2$ wagging, ≈1458 cm$^{-1}$ to a mixture of CH$_2$ scissoring vibration and CH$_3$ antisymmetric deformation, ≈1556 cm$^{-1}$ to asymmetric NH$_3$ deformation, and ≈1584 cm$^{-1}$ to NH$_3^+$ degenerate deformation while the mode at ≈1323 cm$^{-1}$ marked with a blue asterisk is not reported in the literature (Figure S7) [21,24,27,28]. A plausible explanation for the 1323 cm$^{-1}$ mode could be that a strong interaction between BA and MA molecules of RP2 on gold with incoming photons has led to the splitting of mode at ≈ 1320 cm$^{-1}$ into 1300 and 1323 cm$^{-1}$ [29]. This further emphasizes that enhanced light-matter interaction leads to an increase in the intensity of the nominally weak Raman modes of HOIP crystals in addition to observation of hitherto unknown and unseen modes.

## 4. CONCLUSIONS

In summary, we show that the Raman scattering in 2D HOIP crystals can be enhanced without compromising the crystal quality or need for complex integration. To enhance E-field in the crystals and consequently phonon scattering, we fabricated an open-cavity comprising 2D HOIP crystals of varying thicknesses on gold substrates. Due to the Purcell effect we also observe an increase in the Raman signal intensity by a factor up to 15x including observation of new Raman modes at low wavenumbers. TMM simulations corroborate the E-field enhancement hypothesis that we deduce from experiments. Our observations provide a path forward to leverage simple cavity effects induced by a high index sample on a reflective metal substrate to enhance light matter interactions and Raman scattering in materials with phonon scattering cross- sections, and can be potentially translated to other van der Waals crystals also.

**ASSOCIATED CONTENT**

**Supporting Information**

AFM topography and height profiles of RP (n=1-3) crystals; Reflectance of thick and thin RP/gold; Raman spectra of RP2/gold-NPs

**AUTHOR INFORMATION**

**Corresponding Author**

**Deep Jariwala-** [2]*Department of Electrical and Systems Engineering, University of Pennsylvania, Philadelphia, Pennsylvania 19104, United States*
Email: dmj@seas.upenn.edu, Phone: 847-708-4755 (M)/ 215-746-4380 (O)

**Authors**

**Aditya Singh-** [1]*Department of Physics, Indian Institute of Technology Delhi, New Delhi, India-110016;*

[2]*Department of Electrical and Systems Engineering, University of Pennsylvania, Philadelphia, Pennsylvania 19104, United States*

**Jason Lynch-** [2]*Department of Electrical and Systems Engineering, University of Pennsylvania, Philadelphia, Pennsylvania 19104, United States*

**Surendra B. Anantharaman**- [2]*Department of Electrical and Systems Engineering, University of Pennsylvania, Philadelphia, Pennsylvania 19104, United States*

**Jin Hou-** [3]*Department of Chemical and Biomolecular Engineering and Department of Materials Science and Nanoengineering, Rice University, Houston, Texas 77005, United States*

**Simrjit Singh-** [2]*Department of Electrical and Systems Engineering, University of Pennsylvania, Philadelphia, Pennsylvania 19104, United States*

**Gwangwoo Kim-** [2]*Department of Electrical and Systems Engineering, University of Pennsylvania, Philadelphia, Pennsylvania 19104, United States*

**Aditya D. Mohite-** [3]*Department of Chemical and Biomolecular Engineering and Department of Materials Science and Nanoengineering, Rice University, Houston, Texas 77005, United States*

**Rajendra Singh-** [1]*Department of Physics, Indian Institute of Technology Delhi, New Delhi-110016, India*

**Notes**- The authors declare no competing financial interest.

**ACKNOWLEDGMENTS**

D.J. acknowledges primary support for this work by the U.S. Army Research Office under contract number W911NF-19-1-0109. D.J and G.K. acknowledge partial support from Asian office of Aerospace Research and Development (AOARD)/Air Force Office of Scientific Research (AFOSR) grant FA2386-20-1-4074 and partial support from National Science Foundation funded University of Pennsylvania Materials Research Science and Engineering Center (MRSEC) (DMR-1720530). D.J. and J.L. also acknowledge partial support from and FA2386-21-1-4063. A.S. and S.S. acknowledge funding support from the Fulbright-Nehru Fellowship program supported by United State-India Educational Foundation. S.B.A. gratefully acknowledges partial funding received from the Swiss National Science Foundation (SNSF) under the Early Postdoc Mobility program (grant P2ELP2_187977) to conduct this work. The work at Rice University was supported by the US Department of Defense Short-Term Innovative Research (STIR) programme funded by the Army Research Office. R.S. acknowledges partial financial support from the Nanoscale Research Facility at the Indian Institute of Technology Delhi, India.

## TOC Image

*Schematic illustration of enhanced Raman signals from 2D hybrid organic-inorganic perovskites due to open-cavity formation.*

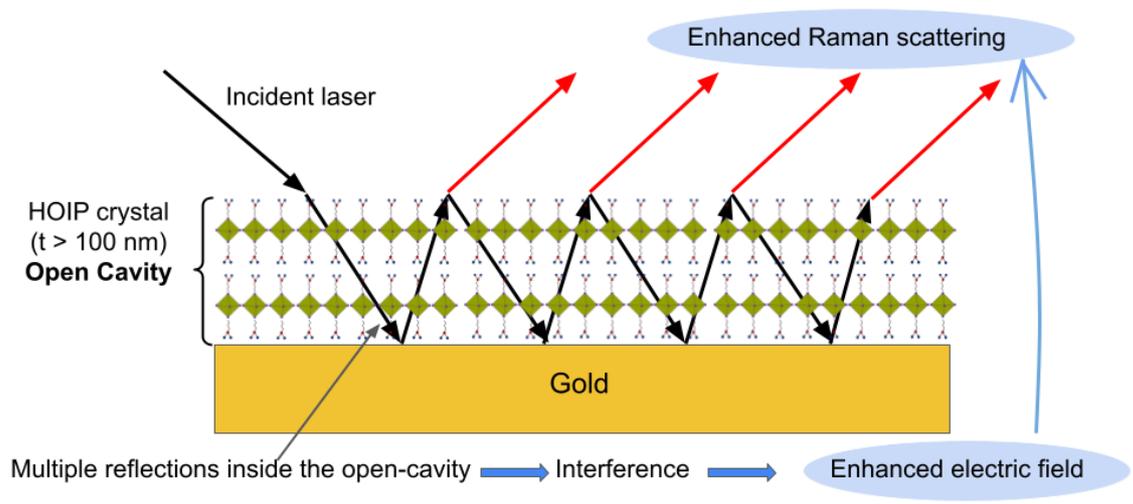

# Cavity-Enhanced Raman Scattering from 2D Hybrid Perovskites

Aditya Singh[1,2], Jason Lynch[2], Surendra B. Anantharaman[2], Jin Hou[3], Simrjit Singh[2], Gwangwoo Kim[2], Aditya D. Mohite[3], Rajendra Singh[1], and Deep Jariwala[2*]

[1]Department of Physics, Indian Institute of Technology Delhi, New Delhi-110016, India
[2]Department of Electrical and Systems Engineering, University of Pennsylvania, Philadelphia, Pennsylvania 19104, United States
[3]Department of Chemical and Biomolecular Engineering and Department of Materials Science and Nanoengineering, Rice University, Houston, Texas 77005, United States

*Corresponding author: dmj@seas.upenn.edu

## S1: Reflectance of 2D RP HOIP thin and thick crystals

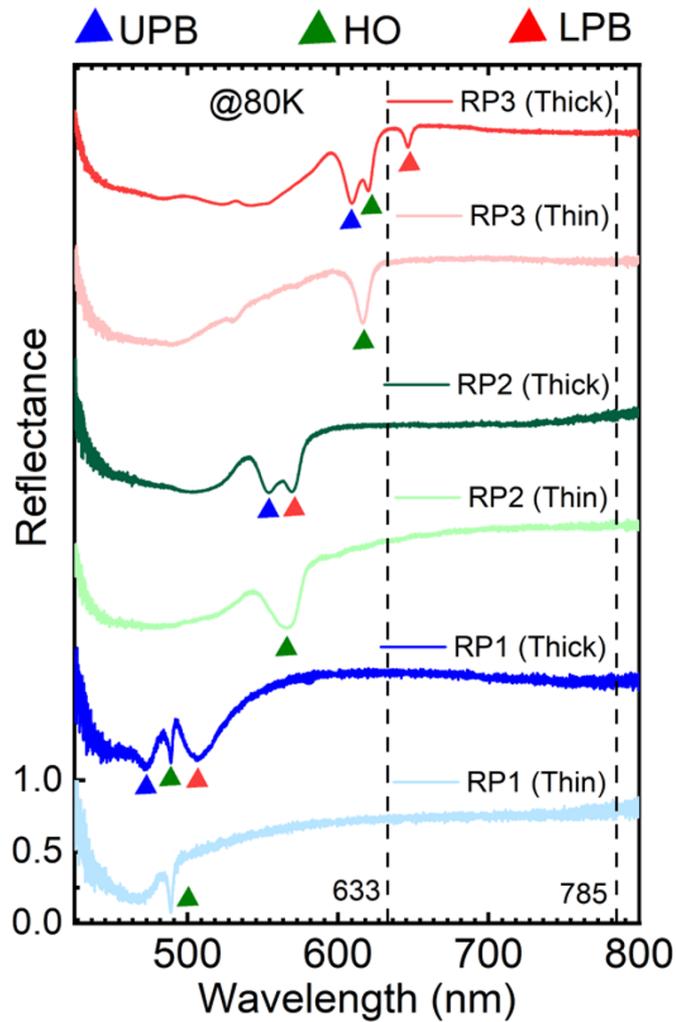

**Figure S1.** Reflectance of thin (thickness t < 30 nm) and thick (t > 100 nm) RP flakes on gold template measured at 80K. As thick flakes forms cavity, polaritonic absorptions corresponding to lower polariton branch (LPB) and upper polariton branch (UPB) are observed in reflectance spectra of thick flakes only, for all three RP. Vertical lines at 633 and 785 nm are drawn for lasers.

## S2. Atomic force microscopy of 2D RP HOIP crystals

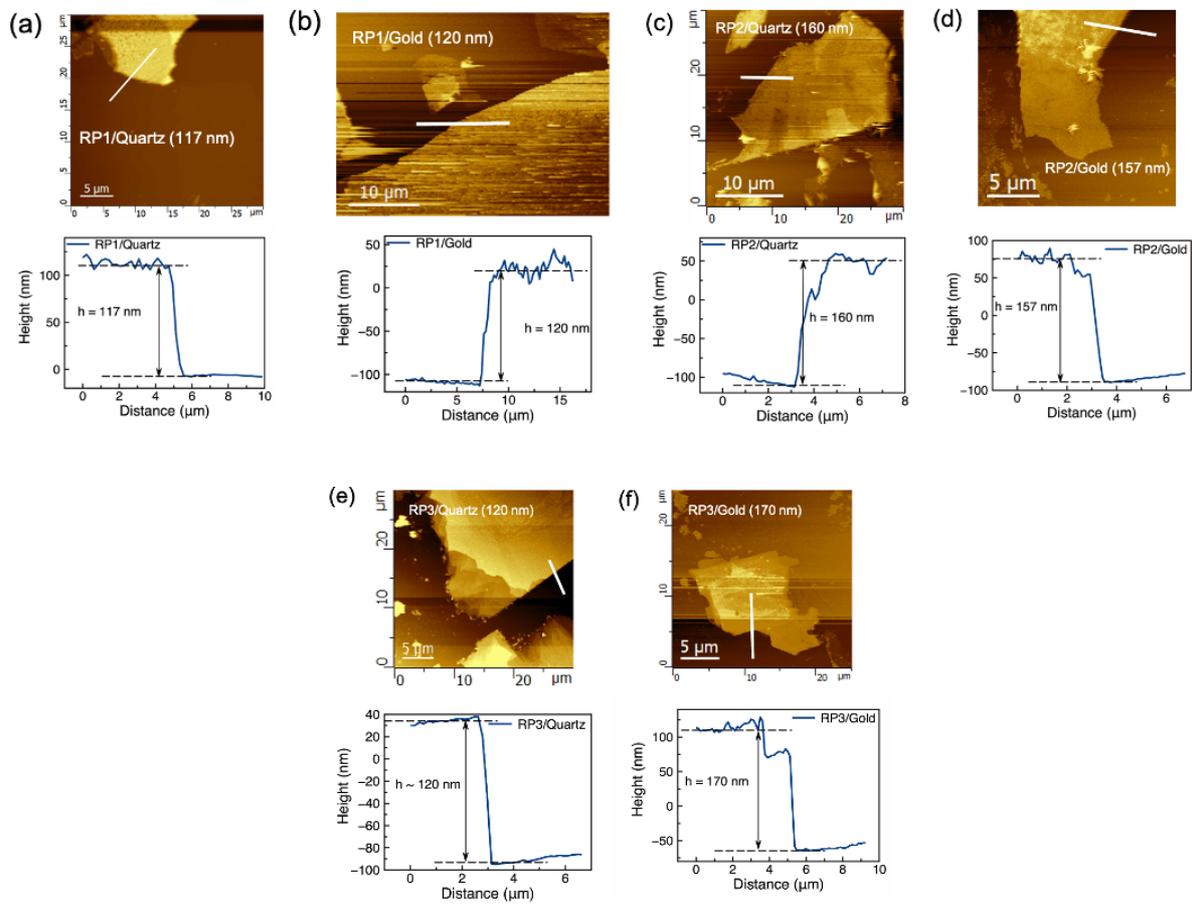

**Figure S2.** Atomic force microscopy (AFM) scan images and height profiles, corresponding to white line drawn on scan images for RPs on gold and quartz.

**S3: Transmission profile of edge filters**

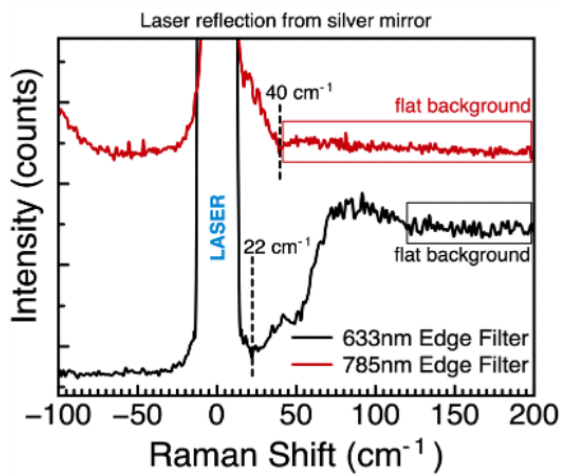

**Figure S3:** Transmission profile of 633 and 785 nm edge filters.

**S4: Comparison of Raman intensity of graphite measured using 633 and 785 nm lasers**

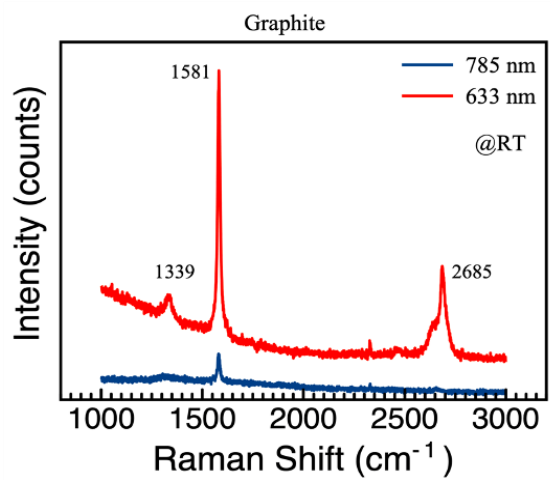

**Figure S4:** Raman spectra of graphite using 785 and 633 nm lasers. Intensity of Raman modes from 785 nm laser are ~ 10x lesser than from the 633 nm laser.

**S5: Raman spectrum of RP3/quartz using 633 nm laser**

Raman spectrum of RP3/quartz was measured using 633 nm laser source at 80K. As RP3 has excitonic absorption at ~ 620 nm, so, 633 nm laser is near-resonant to RP3 HOIP crystal's electronic excitation. Even in near-resonant Raman scattering condition, the Raman modes are not well resolved as compared to RP3/gold (Figures 2a-b). This shows that although 633 nm laser leads to near-resonant Raman scattering, the background emission overshadows the Raman scattering signal.

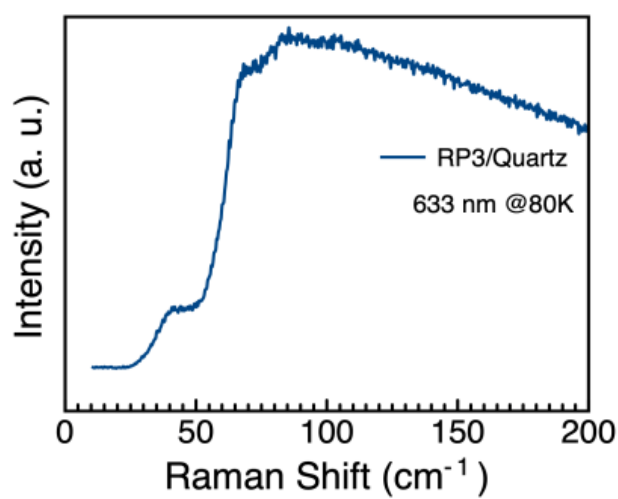

**Figure S5:** Raman spectrum of RP3/quartz obtained using 633 nm excitation laser source at 80K. The thickness of RP3 flake is ~ 120 nm.

**S6: Verification of the role of surface plasmons in Raman scattering**

To find the role of surface plasmons in enhanced Raman signals, we did a control experiment of surface-enhanced Raman spectroscopy (SERS) of RP2 crystal. For that, we used ~13 nm sized gold (Au) nanoparticles (NPs) on silicon substrate, whose plasma oscillation is at 620 nm. RP2 HOIP crystals were exfoliated on gold NPs substrate and as, HOIP crystal are prone to heat, these samples were not heated. Since, the thickness of HOIP flakes is > 100 nm, these crystals are just lying over the NPs and SERS hot spot is away from the HOIP crystal. Figures S6 (d)-(g) clearly show that there is no effect of SERS on RP2/gold-NPs.

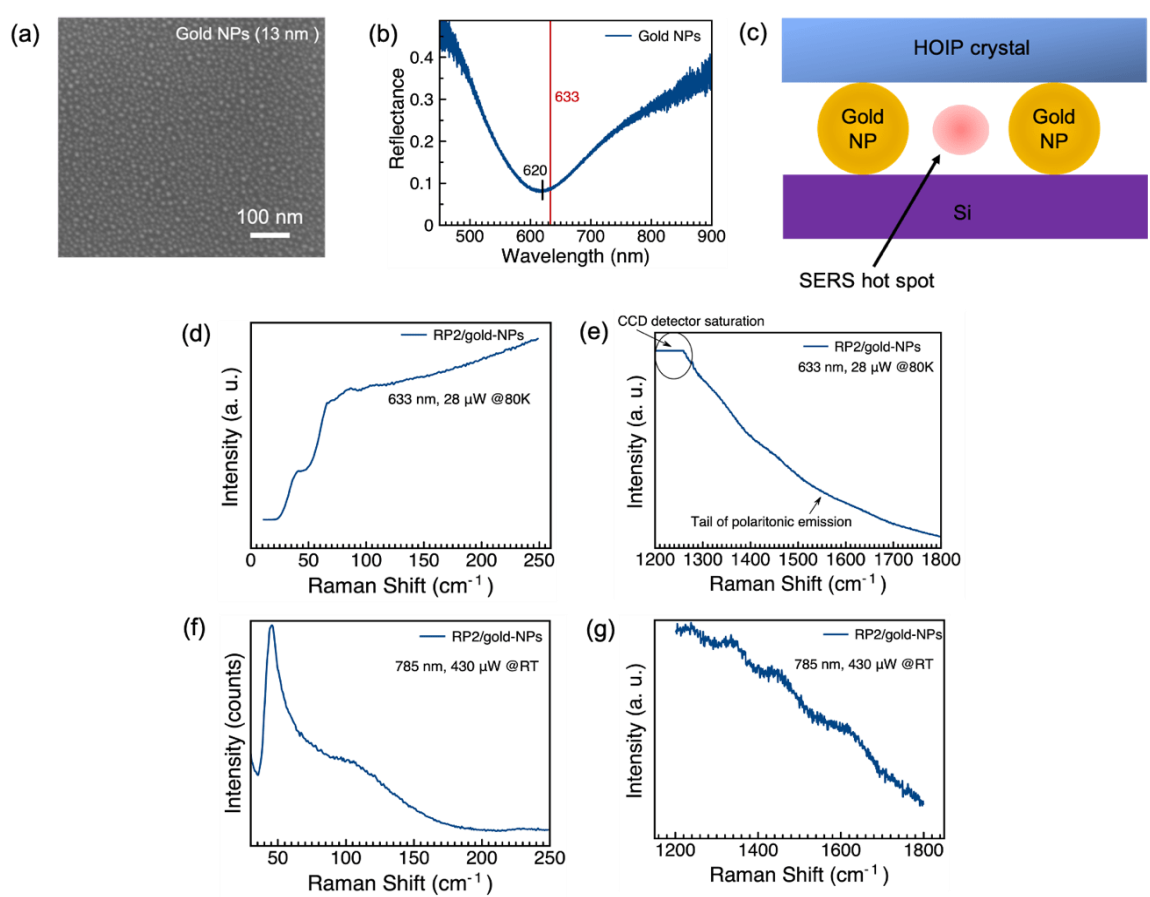

**Figure S6:** (a) Scanning electron microscope image of ~13 nm sized gold nanoparticles (NPs) on silicon substrate. (b) Reflectance of gold NPs which shows plasma oscillation at 620 nm, and 633 nm laser (vertical line in red) is near-resonant to that. (c) Schematic illustration of HOIP crystal on gold NPs substrate depicting that crystal is away from the SERS hot spot.

**S7: Raman modes from organic layer in RP HOIP**

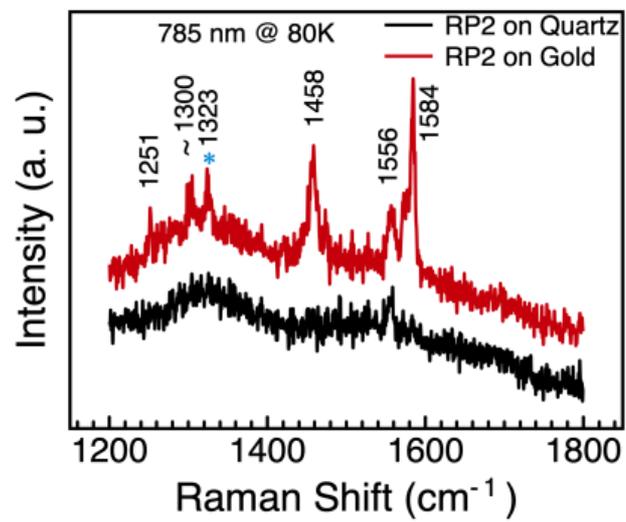

**Figure S7.** Raman modes of RP2 in range R2, 1200-1800 cm$^{-1}$.